 \documentclass[nolinenumbers]{aastex63}

\def\a{\"a}
\def\o{\"o}
\def\u{\"u}
\def\A{\"A}
\def\O{\"O}
\def\U{\"U}

\def\ang{\AA}

\def\gapprox{\lower.4ex\hbox{$\;\buildrel >\over{\scriptstyle\sim}\;$}}
\def\lapprox{\lower.4ex\hbox{$\;\buildrel <\over{\scriptstyle\sim}\;$}}
\def\ref#1{\par\noindent\hangindent1cm {#1}}

\begin{document}

\title{	   The Solar Memory From Hours to Decades	}
 
\correspondingauthor{Markus J. Aschwanden}

\author{Markus J. Aschwanden}
\email{aschwanden@lmsal.com}
\affiliation{Solar and Stellar Astrophysics Laboratory (LMSAL),
 Palo Alto, CA 94304, USA}

\author{Jay R. Johnson}
%\author{Yosia I. Nurhan}
\affiliation{Andrews University, Berrien Springs, MI 49104, USA}

\begin{abstract}
Waiting time distributions allow us to distinguish at least
three different types of dynamical systems, such as (i) linear
random processes (with no memory); (ii) nonlinear, avalanche-type,
nonstationary Poisson processes (with memory during the exponential
growth of the avalanche rise time); and (iii) chaotic systems in the state of
a nonlinear limit cycle (with memory during the oscillatory phase). 
We describe the temporal evolution of the flare rate $\lambda(t) \propto t^p$
with a polynomial function, which allows us to distinguish linear 
($p \approx 1$) from nonlinear ($p \gapprox 2$) events. 
The power law slopes $\alpha$ of observed waiting times (with full solar
cycle coverage) cover a range of $\alpha=2.1-2.4$, which agrees 
well with our prediction of $\alpha = 2.0+1/p = 2.3-2.5$. 
The memory time can also be defined with the time evolution of the
logistic equation, for which we find a relationship between the 
nonlinear growth time $\tau_G = \tau_{rise}/(4p)$ and 
the nonlinearity index $p$. We find a nonlinear evolution for most 
events, in particular for 
the clustering of solar flares ($p=2.2\pm0.1$), 
partially occulted flare events ($p=1.8\pm0.2$), and 
the solar dynamo ($p=2.8\pm0.5$).
The Sun exhibits memory on time scales of 
$\lapprox$2 hours to 3 days (for solar flare clustering),
6 to 23 days (for partially occulted flare events), and
1.5 month to 1 year (for the rise time of the solar dynamo). 

\end{abstract} 


\section{	Introduction			}

The simplest timing information we can obtain from astrophysical data
is probably an event catalog that contains the (start or peak)
times $t_i$ of some phenomenon, such as solar or stellar flares,
observed at some chosen wavelength. An immediate derivation of
this parameter is the so-called waiting time $\tau=t_{i+1}-t_i$
(also called interval time, elapsed time, or laminar time).
From the statistical analysis of such data we can distinguish 
between at least three different dynamical systems: linear
random processes, nonlinear (avalanche) processes, and oscillatory
chaotic systems in the state of limit cycles. The dynamical properties
of these three types of systems is manifested in their waiting
time distributions: (i) linear (stationary) random processes exhibit 
exponentially dropping-off distribution functions; (ii) nonlinear 
(nonstationary Poissonian) processes display power law-like
distribution functions; and (iii) chaotic systems with oscillatory
limit-cycle behavior reveal periodic processes. Another 
distinguishing criterion is their ``memory'' capability:
(i) random processes are incoherent and have no memory 
in consecutive random fluctuations;
(ii) nonlinear (avalanche) events are typically exponentially growing 
and have a memory for the duration of their rise time;
(iii) while (chaotic) limit-cycle behavior has a memory that
lasts at least as long as the oscillatory phase. 
Needless to say that the observational and statistical analysis
of such systems has far-reaching consequences in 
identifying and modeling the underlying physical mechanisms.

Analysis of waiting time distributions and interpretations
in terms of nonlinear system dynamics has been explored
mostly in solar flare data sets (Wheatland et al.~1998;
Boffetta et al.~1999; Wheatland 2000a;
Leddon 2001; Lepreti et al. 2001; Norman et al.~2001;
Wheatland and Litvinenko 2002; Grigolini et al.~2002;
Aschwanden and McTiernan 2010; Gorobets and Messerotti 2012;
Hudson 2020; Morales and Santos 2020).
Similar waiting time distributions have been found in other
solar data sets, such as
in coronal mass ejections (Wheatland 2003; Yeh et al.~2005;
Wang et al.~2013, 2017);
in solar energetic particle events (Li et al.~2014);
in solar wind discontinuities and its intermittent turbulence
(Greco et al.~2009; Wanliss and Weygand 2007),
in heliospheric type III radio burst storms
(Eastwood et al.~2010), and
in solar wind switchback events (Bourouaine et al.~2020;
Dudok de Wit et al.~2020; Aschwanden and Dudok de Wit 2021).
The cyclic behavior of the solar dynamo has been established
over millennia (Usoskin et al.~2017).
Extending out to stars, waiting time distributions were studied
in active and inactive M-dwarf stars (Hawley et al.~2014;
Li et al.~2018),
in the avalanche dynamics of radio pulsar glitches and in 
gamma-ray bursts (Guidorzi et al.~2015; Yi et al.~2016), and
black hole systems (Wang et al.~2015, 2017).

It was recognized that a deeper understanding of waiting
time distributions requires a physically motivated model
of the variability of the flare rate function. However,
the power law slope $\alpha$ of waiting time distributions,  
$N(\tau) \propto \tau^{-\alpha}$, does not exhibit
a unique value, but is found to vary in a range of
$1 \lapprox \alpha \lapprox 3$ (Wheatland and Litvinenko 2002;
Aschwanden and McTiernan 2010). It was noted that the
variability of the flare rate strongly depends on the
phase of the solar cycle (Aschwanden and Dudok de Wit 2021;
Aschwanden, Johnson, and Nurhan 2021). 

In this study we focus on the ``memory'' of solar processes,
which we characterize with the duration of coherent growth
in solar time structures. We model the time evolution
of the event rate during the initial exponential growth 
phase with a polynomial function, $\lambda(t) \propto t^p$. 
This coherent growth phase is also typical for 
avalanching events that occur in self-organized criticality 
models (Aschwanden 2011).  
The time evolution of avalanching flare rates can also
be described with the logistic first-order differential
equation (Aschwanden et al.~1998, Wang et al.~2009; 
Aschwanden 2012b; Qin and Wu 2018), which is almost
identical to the polynomial model (see comparisons in Fig.~1).
We measure the degree of nonlinearity, which controls
the coherent evolution during the rise time of an
instability, and find that the Sun
has a memory over time scales varying by at least four
orders of magnitude, from clustering of solar flares 
(on time scales of $\lapprox 2$ hours) to the 
dynamo-driven solar cycle (on time scales of several decades). 

The content of this paper includes a brief description of the
theory (Section 2), data analysis of GOES data (Section 3), 
a discussion of previous work (Section 4), and conclusions 
(Section 5). 
 
\section{	Theory 		 }

\subsection{		The Waiting Time Distribution 	}

Waiting time distributions, $N(\tau) \ d\tau$, have the diagnostic
potential to reveal whether solar flares are generated by a
stochastic (or random) stationary Poisson process (if they obey
an exponential distribution, $N(\tau) \propto \exp[-\tau]$),
or by a nonstationary (nonlinear) Poission process (if they obey   
a power law-like distribution, $N(\tau) \propto \tau^{-\alpha}$) 
(Wheatland and Litvinenko 2002). 
The difference between a stationary and a nonstationary Poisson
process is generally quantified by the statistical behavior 
of the flare rate function $\Lambda(t)$, which can be constant,
i.e., $\Lambda(t)=\Lambda_0$, at one extreme, or can be
highly time-variable, following an arbitrary temporal function
$\Lambda(t)$, in the other extreme. Observations generally
exhibit nonstationary flare rate functions, $\Lambda(t) \neq 
const$ (Wheatland et al.~1998; Wheatland 2000c; 2006;
Wheatland and Litvinenko 2002; 
Aschwanden and McTiernan 2010; Aschwanden 2019b).
While we denote the entirely observed time profile of the flare rate
with the symbol $\Lambda(t)$, time segments of coherent
growth are denoted with the symbol $\lambda(t)$.
Various analytical incarnations of the flare rate function 
$\lambda(t)$ have been used to calculate waiting time
distributions, such as 
a polynomial function $\lambda(t) \propto t^p$,
a sinusoidal function $\lambda(t) \propto \sin(t)^p$, or 
a gaussian function $\lambda(t) \propto \exp(-t^2)$, 
for which exact analytical solutions were found recently,
in terms of Bessel functions (Nurhan et al.~2021)
and the incomplete gamma function (Aschwanden, 
Johnson, and Nurhan 2021). 

The power law slope $\alpha$ of waiting time distributions,  
$N(\tau) \propto \tau^{-\alpha}$, does not exhibit
a unique value, but is found to vary in a range of
$1 \lapprox \alpha \lapprox 3$ (Wheatland and Litvinenko 2002;
Aschwanden and McTiernan 2010). This variable behavior was 
attributed to intrinsically different flare rate evolutions 
$\lambda(t)$ during the solar cycle minimum and maximum phase 
(Wheatland and Litvinenko 2002). More specifically, recent
work has shown that the power law slope $\alpha$ of waiting
time distributions depends on the nonlinearity index $p(\alpha)$
of the polynomial flare rate evolution $\lambda(t) \propto t^p$ 
in a unique way (Aschwanden, Johnson, and Nurhan 2021),
\begin{equation}
	\alpha = 2 + {1 \over p} \ ,
\end{equation}
based on the calculation of the exact analytical solution
in terms of the incomplete Gamma function $\gamma[\alpha, \beta]$,
with $\alpha$ the power law slope of the waiting time distribution, and 
with the argument $\beta = \lambda_0 \tau$, where $\lambda_0$ is the mean
flare rate and $\tau$ is the waiting time,
\begin{equation}
	N(\tau, \alpha)\ d\tau = \lambda_0 (\alpha - 1)
	\ \gamma[\alpha, \beta = \lambda_0 \tau]
	\ (\lambda_0 \tau)^{-\alpha} \ d\tau \ .
\end{equation}

A practical approximation is an expression in terms of the
Gamma function $\Gamma[\alpha]$,
\begin{equation}
	N(\tau, \alpha)\ d\tau = \lambda_0 (\alpha - 1)
	\ \Gamma[\alpha]
	\ (\lambda_0 \tau)^{-\alpha} \ d\tau \ ,
\end{equation}
which leads to a straight power law, $N(\tau) \propto 
\tau^{-\alpha}$, and agrees with the exact analytical
solution in the asymptotic limit of large waiting times.
Note that the definition of the Gamma function entails
an integral with the limits $[0,\infty]$, while the
incomplete Gamma function has finite integral limits
$[0,\beta]$ (see Eqs.~(20) and (24) in Aschwanden et al.~2021),
and applies in the asymptotic regime $\lambda_0 \tau \gg 1$
where a power law function is found.

\subsection{		The Polynomial Flare Rate Function		}

As we will see in the following, waiting time distributions
can only be understood with the knowledge of the temporal
variability of the flare rate.
The variability of the flare rate can most easily be 
quantified by histograms. Such a histogram
of the flare rate $\Lambda(t)$ is shown in Fig.~2a, with four 
different time resolutions, expressed by the 
time bin widths $\Delta t$ = (1 year)/$n_{bin}$ for 
$n_{bin}=2$ (Fig.~2a),
$n_{bin}=8$ (Fig.~2b), 
$n_{bin}=32$ (Fig.~2c), and 
$n_{bin}=128$ (Fig.~2d). 
The histograms are sampled from the entire data set of 338,661 flare 
events during 37 years (1974-2012). The time profile $\Lambda(t)$ shown
in Fig.~2 reveals four major time structures at low time resolution (Fig.~2a),
each one encompassing a time duration of $\approx 11$ years, 
which obviously are attributed to the magnetic 
(Hale) solar cycle (Fig.~2a). The shorter time structures, visible at
higher time resolution (Figs.~2c, and 2d), contain
either random noise, nonlinear clustering of flares (indicating
a ``solar memory''), or a combination of both, which is one of 
the main tasks pursued in this study.

In a next step we define time structures $\lambda(t)$.  
They represent partial time segments and are extracted from the entire 
time profile $\Lambda(t)$ extending over the total (37-year) time 
interval of the observed flare rates. 
A first selection criterion of time structures is made
by requiring a local peak $\lambda(t=t_{max})$ in the time profile 
$\Lambda(t)$,
\begin{equation}
	\lambda(t_{i-1}) < \lambda(t_i) = \lambda(t=t_{max}) > \lambda(t_{i+1}) \ .
\end{equation}
Secondly, we define the absolute minimum $t_{min}$ 
between two subsequent peaks,
\begin{equation}
	\lambda(t_{max,j-1}) > \lambda(t=t_{min}) < \lambda(t_{max,j+1}) \ .
\end{equation}
The time interval between a minimum at $t_{min}$ and a maximum
at $t_{max}$ is characterized by a monotonic increase in the flare rate.
The simplest definition of a time structure would be a 
linear segment from $\lambda_{min}=\lambda(t=t_{min})$ to $\lambda_{max}=\lambda(t=t_{max})$. 
However, in order to make our time structures capable to distinguish
between linear and nonlinear time structures, we generalize the
linear exponent $p=1$ to a nonlinearity index $p \ge 1$,
\begin{equation}
	\lambda(t) = \left\{ 
	\begin{array}{ll}
	\lambda_1 + (\lambda_0-\lambda_1) \left[ {(t - t_1)/(t_0 - t_1)} \right]^p 
		& \quad{\rm for}\ t_1 \le t \le t_0 \\
	\lambda_2 + (\lambda_0-\lambda_2) \left[ {(t_2 - t ) / (t_2 - t_0)} \right]^p 
		& \quad{\rm for}\ t_0 \le t \le t_2 \\
	\end{array}
	\right. \ .
\end{equation}
A graphical definition of such a rise time structure is shown in Fig.~1,
which covers a time interval of $[t_1,t_2]=[t_{min},t_{max}]$, and has an 
inflection point at $[t_0, \lambda_0]$. Note that the time profile
in the time range $t_1 \le t \le t_2$ before the inflection point
at $t_0$ is identical with the polynomial definition in
Aschwanden et al.~(2021). Such a time structure is defined 
by 7 parameters: by four constants $[t_1,t_2,\lambda_1,\lambda_2]$ and three free
parameters $[t_0,\lambda_0,p]$, where $[t_0,\lambda_0]$ define the inflection point
and $p$ is the nonlinearity index. Since there are 3 free parameters
we require at least 4 or 5 time bins per fitted time structure.

This definition of time structures has the capability to discriminate
between linear $(p=1)$ and nonlinear $(p \gapprox 2)$ time evolutions.
Furthermore, it reveals time structures in a large range of time
resolutions, but is far from complete event detection, especially
for noisy structures with a duration of less than 4 to 5 time bins.
Thus, this event detection algorithm is biased towards long-duration
time structures, but should not be biased with respect to linear
versus nonlinear event statistics. 

\subsection{		The Logistic Flare Rate Function		}

In the previous section we defined the polynomial flare rate function
$\lambda(t)$, which provides us a diagnostic whether the flare rate
is stochastic, without any memory (if $p \approx 1$), 
or has some memory (if $p \gapprox 2$). A time structure is defined
here by a time interval with coherent growth in the event rate.
Our parameterization in terms
of a polynomial index $p$, i.e., $\lambda(t) \propto t^p$, was chosen
mostly for reasons of mathematical convenience, and has been  
used in a previous publication (Aschwanden, Johnson, and Nurhan 2021). 
Alternatively, we find that the time evolution of the flare rate can be 
represented with a physical model of logistic growth and saturation, 
which moreover is almost indistinguishable from the polynomial model 
of the flare rate evolution (Fig.~1). 

The time evolution of the logistic growth model (Aschwanden 2011, p.94) 
is universally similar for many instabilities, consisting of an
initial exponential growth phase with subsequent saturation, which
can be described by a simple first-order differential equation
(discovered by Pierre Fran\c{c}ois Verhulst in 1845; May 1974;
Beltrami 1987, Jackson 1989, Aschwanden 2011, p.94), 
\begin{equation}
	{d\lambda(t) \over dt} = \left({\lambda(t) \over \tau_G}\right)
	\left( 1 - {\lambda(t) \over \lambda_{\infty}} \right) \ ,
\end{equation}
where $\lambda(t)$ is the time-dependent event rate here, 
$\lambda_{\infty}$ is the maximum rate asymptotically 
reached at infinite time $t \mapsto \infty$
(also called the {\sl carrying capacity} in ecological applications),
and $\tau_G$ is the e-folding (exponential) growth time. 
In our application here, $\lambda(t)$ is the time-dependent
event rate that monotonically increases during the rise time
of an instability, which is a phase of coherent growth and defines
the start $t_1$ and inflection time $t_0$ 
of an avalanche or cluster of events (Fig.~1).
One can easily devise the evolutionary solution from the 
logistic equation: For small
times we have exponential growth, $d\lambda/dt \approx \lambda(t)/\tau_G$,
while for large times we have progressive saturation 
according to the rightmost term, $d\lambda/dt \propto
[1-\lambda(t)/\lambda_\infty]$. The exact solution of this first-order
differential equation is,
\begin{equation}
	\lambda(t) = {\lambda_\infty \over 1 + \exp{ \left( {-(t-t_0) \over \tau_G} \right)}} \ ,
\end{equation}
with an inflection point at $[t_0, \lambda_0]$,
\begin{equation}
	\lambda_0 = \lambda(t=t_0) = {\lambda_{\infty} \over 2} \ .
\end{equation}
Comparing with the polynomial flare rate function (Eq.~6), 
\begin{equation}
	\lambda(t) = \lambda_1 + (\lambda_0-\lambda_1) 
	\left({ t-t_1 \over t_0-t_1}\right)^p \ , \qquad
	t_1 \le t \le t_0 \ ,
\end{equation}
and its time derivative,
\begin{equation}
	{d\lambda(t) \over dt} = p \left({\lambda_0 - \lambda_1 
	\over t_0 - t_1}\right)
	\left( { t - t_1 \over t_0 - t_1 } \right)^{p-1} \ , \qquad 
	t_1 \le t \le t_0 \ ,
\end{equation}
we can equate the values at the inflection point, 
$\lambda_0=\lambda(t=t_0)$ (Eq.~8 and Eq.~10),
as well as their time derivatives $d\lambda(t=t_0)/dt$ (Eq.~7 and Eq.~11),
for the polynomial and the logistic function. From these equated quantities
we obtain an expression for the relationship of the growth time $\tau_G$ on the
rise time $t_{rise}$ and the nonlinearity index $p$,
\begin{equation}
	\tau_G = {1 \over 2 p} \left( {(t_0-t_1) \over 1 - \lambda_1/\lambda_0} \right) 
	\approx  {(t_0-t_1) \over 2 p} 
	= {\tau_{rise} \over 4 p}
\end{equation}
where the rise time is defined by $\tau_{rise}=(t_2-t_1)=2(t_0-t_1)$ (see Fig.~1).
The right-hand expression is an approximation based on 
$\lambda_1 \ll \lambda_0 = \lambda_\infty/2$ for $t \mapsto \infty$. 
	
The application of the logistic equation to the flare rate function
$\lambda(t)$ here implies a well-defined time structure that consists 
of a coherent growth phase and a subsequent saturation phase, and thus 
exhibits memory during this rise time interval $\tau_{rise}$.
This approach is mathematically convenient, has a physical meaning,
and is quantified with a very simple expression to the nonlinearity
factor $p$, namely $\tau_G \approx \tau_{rise}/4 p$. 

\section{	Data Analysis 	  }

\subsection{		The Enhanced GOES Flare Catalog	}

For solar flare events, 
an official flare catalog is issued by the {\sl National 
Oceanographic and Atmospheric Administration (NOAA)} 
(http://www.oso.noaa.gov/goes/), based on observations with the 
{\sl Geostationary Operational Environmental Satellites (GOES)}, 
which covers now 47 years of observations (1974-2021).

For the purpose of statistsical analysis it is generally reommended
to use the largest available data sets. 
The largest solar flare catalog has been created with an automated flare
detection algorithm, applied to the GOES 1-8 \ang\ light curves,
which detected 338,661 flare events during 37 years (1974-2011)
(Aschwanden and Freeland 2012). For comparison, a subset observed
with GOES during 1991-2011 yields 39,696 solar flare events. This 
implies that the automated flare detection algorithm has a 
$\approx 5$ times higher sensitivity than the NOAA flare 
catalog. The flare detection scheme is based on detection of soft X-ray 
flux minima and maxima, after appropriate background subtraction,
thresholding, and data gap elimination. The product is then 
a list of flare peak times, $t_i=1,...,n_{ev}$, detected from
the soft X-ray light curve, which was sampled from a time 
resolution of $\Delta t=12$ s (after rebinning from the original 
$\Delta t=3$ s GOES time resolution). 

A main product used in this analysis here is the statistics 
of waiting times $\tau$, which are simply measured from the 
time intervals of the time-ordered flare peak times $t_i$,
\begin{equation}
	\tau_i = (t_{i+1}-t_i) \ , \quad i=1,...,n_{ev}-1 \ .
\end{equation}
The longest waiting times identified in the enhanced GOES flare
catalog extend up to time scales of months, for instance during
the solar minimum of 2008-2009. Other long
waiting times occur due to the instrumental duty cycle
(varying from 76\% to 94\% per year), due to unreadable data files,
missing data, data loss, telemetry gaps, calibration procedures,
or Earth occultation. Nevertheless, GOES has the
highest duty cycle among all solar-dedicated space missions, and
thus offers the most complete record of solar flare waiting times.

\subsection{	Fitting of the Flare Rate Function	}

The time structure $\lambda(t)$ (Eq.~6) can be fitted to the observed data
$\lambda_{obs}(t)$ by a standard least-square optimization algorithm,
which we use from the {\sl Interactive Data Language (IDL)} software,
\begin{equation}
        \chi =
        \sqrt{ {1 \over (n_{bin} - n_{par})}
        \sum_{i=1}^{n_{bin}}
        {[\lambda(t_i)-\lambda(t_{i,obs})]^2
        \over \sigma_i(t_i)^2 }
        } \ ,
\end{equation}
where $\lambda_i=\lambda(t=t_i),\ i=1,...,n_{bin}$ are the flare rates per bin,
defined by the model given in Eq.~(6),
$\lambda(t_{i,obs})$ are the corresponding obvserved values,
$n_{bin} \ge 4$ is the number of fitted histogram bins,
and $n_{par}=3$ is the number of free parameters
of the fitted model function $\lambda(t)$.
The estimated uncertainty of flare rates per bin,
$\sigma_i$, is according to Poisson statistics,
\begin{equation}
        \sigma_i = {\sqrt{\lambda_{i,obs} \Delta t} \over \Delta t} \ .
\end{equation}
Four examples of fitted time structures are shown in Fig.~2a
(red curves), for the four time structures that are produced
by four solar cycles.

It has been pointed out that a linear regression fit on a 
log-log scale is biased
and inaccurate, while using a maximum likelihood estimation
is more robust (Goldstein et al.~2004; Newman 2005; Bauke 2007).
We use Poissonian weighting (Eq.~15), which theoretically improves
the formal error, but there is a larger systematic error due to 
deviations from ideal power laws, which can only be quantified
by calculating the exact analytical solutions of waiting time
distributions (Aschwanden, Johnson, and Nurhan 2021).

We sampled time structures by using an automated detection
algorithm, for 12 different time resolutions $\Delta t$, 
logarithmically spaced with $n_{bin}=2^i,\ i=0,1,...,12$ per year.
The total of detected structures amounts to 848 time structures,
ranging from time resolutions of $\Delta t=2$ hrs to
$\Delta t=1$ year (Table 1). Each time structure 
was fitted with the polynomial time profile model
$\lambda(t)$ (Eq.~6), and the best-fit three parameters
$[t_0,\lambda_0,p]$ were determined. 
We show the detailed fits from a selection
of 12 events (out of the 848 detected events) in Fig.~3, 
selected from 12 different time
resolutions and cases with the largest number of time bins 
(monotonically increasing during the rise time). For instance,
the first example shown in Fig.~3a is gathered from
$n_{bin}=5$ time bins, a time resolution of 
$\Delta t$ = (1 year)/$n_{bin}=1.0$ year,
a duration of $D=n_{bin} \Delta t=5.0$ years, 
a nonlinearity index of $p=3.4$, and a goodness-of-fit
$\chi=7.0$. Note that the formal error (Eq.~15) 
is an adequate estimate of the statistical uncertainty
in cases with $\chi \lapprox 2$, while high values
of $\chi \gapprox 2$ (e.g., Fig.~3a, 3b) indicate 
under-estimated uncertainties $\chi$ due to very high 
flare rates (of $\lambda(t_{max}) \gapprox 10^4$ years$^{-1}$) and
unknown systematic errors of the model (Eq.~6).

\subsection{	Measurement of The Nonlinearity Index		}

The major new result of this analysis is the measurement
of the nonlinearity index $p$, which represents the order
(or degree) of the polynomial flare rate evolution,
$\lambda(t) \propto t^p$ (Eq.~6). The physical implication 
is that we have a diagnostic whether the evolution
of an event is linear, $\lambda(t) \propto t$, which is
typical for linear random processes, or if the
event evolution is nonlinear, $\lambda(t) \propto t^p$,
which is typical for exponentially growing avalanche
processes.

We list the obtained nonlinearity indices $p$ in Table 1, 
and show their dependence on the time resolution 
$\Delta t = [1\ year]/n_{bin}$ 
in Fig.~4. Interestingly, the graph shown in Fig.~(4) reveals
three different regimes: (i) One group with a cubic nonlinearity
of $p=2.83\pm0.49$ during time ranges from 1.5 months to 1 year,
is evidently produced by the solar dynamo, due to the 11-year
time scale; (ii) Another group with a quadratic nonlinearity of
$p=2.19\pm0.07$ is found during a time range from 2 hours to 3 days,
which is attributed to clustering of solar flares;
and (iii) An intermediate group with a nonlinearity 
of $p=1.83\pm0.22$ during time ranges from 6 days to 23 days,
is most likely affected by the solar rotation rate (which has
a sidereal rotation rate of $\approx 26$ days). 

A remarkable result is that all fitted nonlinearity indices $p$
yield values in a range of $p \approx$ 1.6-3.5 (Fig.~4). This means
that all groups are significantly above the linear (random) range
($p \approx 1$), which suggests that nonlinear physical processes 
are responsible for all detected time structures, for solar flares,
partially occulted long-duration flares, as well as for the solar dynamo. 
These results
on the absence of linear random processes and the ubiquity of
nonlinearity ($p \gapprox 2$) demands a theoretical explanation.

\section{	Discussion				}

The behavior of the nonlinearity index $p$ gives us some deeper
understanding on the waiting time distribution (Section 4.1), 
the solar rotation effect (Section 4.2),
the solar dynamo (Section 4.3), implications for
self-organized criticality models (Section 4.4), and the stochasticity, 
intermittency, and memory of solar flare rates (Section 4.5).

\subsection{	Waiting Time Distributions 	}

From numerical simulations and analytical calculations
of nonstationary waiting time distributions we learned
that their power law slope $\alpha$ depends on the
nonlinearity index $p$, namely $\alpha=2+1/p$ (Eq.~1),
as derived in recent work (Aschwanden, Johnson, and Nurham 2021).
Once the nonlinearity index $p$ is known for a given data set,
the waiting time distribution $N(\tau)$ is in principle
fully determined (with Eqs.~2 or 3), using the analytically derived
relationship (Eq.~1). However, the various studies on the
waiting time distributions $N(\tau)$ with exact analytical
solutions have revealed a high sensitivity on the long waiting
times, which corresponds to the time intervals of minimum flare rates
and occur at the beginning of exponentially growing flare rates,
as modeled with the logistic and polynomial models.
It is instructive to study the differences early in flare
events (say in the range of $t \lapprox 0.2$ in Fig.~1),
which shows relatively large differences of $\approx 2\%$
for a nonlinearity index of $p=2$ (Fig.~1, left), but reveals
much smaller differences of $\approx 0.5\%$ for $p=3$ 
(Fig.~1, right). Thus, this comparison suggests that the 
logistic approximation is generally more accurate (than
the polynomial model) for high nonlinearity indices 
($p \gapprox 3$).
  
A careful analysis of the waiting time distribution for 
the logistic equation shows that a power law typically only
occurs when there is a large increase in rate 
($\lambda_1 / \lambda_2 \gg 0.1$), and for $1 \sim \lambda_1 \Delta \lapprox 10$.
For these parameters, the power law is typically in the range 
$-2 < \alpha < -2.5$ consistent with the power
laws obtained from the approximate fit 
(Eq.~10 with $p \ge 2$) of the solution of the logistic equation. 

Here we find different results of the power law slope $\alpha$
among three different groups in Fig.~4, namely for the
solar dynamo, solar flare clusters, and partially occulted
flares (an effect caused by the solar rotation).
In the following we average the power law slopes from
12 different time resolutions.
Using the results obtained from the solar dynamo,
$p=2.83\pm0.49$ (Fig.~4), we predict a power law slope of 
$\alpha=2+1/p \approx 2.4$ (using Eq.~1).
For solar flares, with $p=2.19\pm0.07$ (Fig.~4), we predict
a power law slope of $\alpha=2+1/p \approx 2.5$.
For the intermediate group, which is affected by the solar
rotation with $p=1.83\pm0.22$ (Fig.~4), we predict a power law
slope of $\alpha=2+1/p \approx 2.5$. Thus these three 
cases cover a range of $\alpha \approx 2.4-2.5$.
This result indeed matches closely the power law slopes
observed from previous GOES waiting time distributions
in the range of $\alpha \approx 2.1-2.4$, reported as
$\alpha=2.4\pm0.1$ (Boffetta et al.~1999),
$\alpha=2.16\pm0.05$ (Wheatland 2000a; Lepreti et al.~2001);
$\alpha=2.36\pm0.11$ (Wheatland and Litvinenko 2002).
A similar value of $\alpha=2.5$ was recently calculated
for a sinusoidal flare rate function also (Nurhan et al.~2021),
instead of the polynomial flare rate model used here.

\subsection{		Solar Rotation Effect	}

The temporal variation of the solar flare rate has been studied
in terms of the fractal dimension in the case of solar radio
emission (Watari 1996a), which led to a diagnostic for 
periodic, chaotic, and random components (Watari 1996b).
A power spectrum $<P(\tau)>$ of the daily sunspot number and
radio flux has been calculated, with $\tau$ a time interval, 
yielding a fractal relationship $<P(\tau)> \propto \tau^{-\alpha}$
and a power law slope in the range of $\alpha \approx 1.2-2.0$.
Interestingly they find an effect of the solar rotation,
which they simulate with and without the rotational effect.
They find a steepening of the power spectrum slope at
a time interval of $\tau \gapprox 26$ days.

The manifestation of a solar rotation effect simulated in
Watari (1996a, 1996b) affects our data analysis similarly. 
We carried out some simplified modeling and found that
long waiting times are over-represented due to occulting
at the solar limb. Solar occulting artificially increases
the number of waiting times at $\tau = T_{rot}/2$, as there is
an artificial bias that introduces a spurious excess of
waiting times at the rotation period.

\subsection{		The Solar Dynamo	}

The solar dynamo reverses the solar magnetic field every 11 years,
which yields a 22-year cycle for the same magnetic polarity, also
called the Hale magnetic cycle. The observational manifestation
of this cyclic behavior is also reflected in the decadal variability
of the flare rate $\Lambda(t)$ and the power law slope of flare durations
(Aschwanden and Freeland 2012), as well as in the variability of
the sunspot number (for a recent analysis see Aschwanden and 
Dudok de Wit 2021). The underlying physical mechanism
is a quasi-stationary oscillation of a nonlinear system that is
called the {\sl limit cycle} (for a miniature-review see chapter 3.6
in Aschwanden 2011). It occurs in many nonlinear systems
that come close to an oscillatory behavior. Theoretical examples
of such nonlinear systems are the Hopf bifurcation (Cameron
and Schuessler 2017), or the Lotka-Volterra coupled differential
equation system (Consolini et al.~2009) known in ecological sciences
(May 1974). The physical mechanism of the solar dynamo cycle can ultimately
be understood as a near-equilibrium oscillation between the global
solar poloidal magnetic field $B_r(t)$ and the toroidal magnetic field
$B_\theta(t)$ (Charbonneau 2005; Cameron and Schuessler 2017). 

Regarding our polynomial approach of characterizing the flare rate
variability, $\lambda(r) \propto t^p$, the nonlinearity parameter $p$ is
an observable, for which values of $p=2.83\pm0.49$ (Table 1 and
Fig.~4) were found. Alternatively, instead of assuming a general
polynomial index $p$, a sinusoidal model $\lambda(\tau) \propto 
1+\cos(\tau)$ has been used also (Nurhan et al.~2021), yielding 
a power law-like waiting distribution with a slope of $\alpha=2.5$.
Based on the predicted relationship $\alpha=2+1/p$ (Eq.~1) we expect
to measure a nonlinearity index of $p=1/(\alpha-2)=2$, 
which indeed confirms the expected nonlinearity range of $p \gapprox 2$.
Some differences could possibly be explained with the asymmetry
of the solar cycle time profile $\Lambda(t)$. More specifically, 
the rise time of a sunspot cycle varies inversely with the cycle 
amplitude: Strong cycles rise to their maximum faster than weak cycles,
also known as {\sl Waldmeier effect}. 

\subsection{		Self-Organized Criticality Models	}

The most conspicuous feature of {\sl self-organized criticality (SOC)}
models is the avalanche behavior of events in a nonlinear dissipative system,
leading to power law slopes of their size distribution and their
duration distribution (Bak et al.~1987; 1988). The simplest dynamic
model of a SOC avalanche can be described by an (initial) exponential
growth phase after the onset of an instability, and saturation of
the instability after a random time. These two assumptions directly
predict a power law-like distribution of avalanche sizes.
Besides this exponential-growth model (Willis and Yule 1922;
Rosner and Vaiana 1978; Aschwanden 1998), a power law-growth model
(which is equivalent to our polynomial time evolution), and a
logistic-growth model have also been formulated (Aschwanden 2011).
(e.g., May 1974; Beltrami 1987, Jackson 1989, Aschwanden 2011, p.94), 

If we interpret time structures with a nonlinearity index $p$ as
SOC avalanches, we need to test whether a polynomial event rate model
$\lambda(t) \propto t^p$ is consistent with a logistic model.
Such a comparison is shown in Fig.~1, where the two functions 
are almost indistinguishable.
Consequently, we can use the logistic 
model and the polynomial model equally well as a discriminative 
diagnostic between linear ($p\approx 1$) and nonlinear 
($p \gapprox 2$) systems. This has far-reaching consequences
for SOC models. Traditional SOC models assume a slow-driven
and stationary flaring rate (Bak et al.~1987, 1988), while
more recent studies adjust to (i) multiple energy dissipation
episodes during individual flares, (ii) violation of 
time scale separation (between flare durations and waiting times),
and (iii) fast-driven and nonstationary flaring rates
(Aschwanden 2019b). 

\subsection{		Stochasticity and Memory	}

There are at least three unmistakable dynamic patterns of 
dissipative systems: linear, non-linear, and limit-cycle systems. 

The first mechanism is a linear system, where the total dissipated
energy grows linearly with the energy input, the energy dissipation
rate or event rate is constant, and the resulting waiting
time distribution is exponentially dropping off, $N(\tau) \propto
\exp{[-\tau]}$, according to a random process. As an example 
we consider the
accumulated number of photons emitted from the Sun or a star in
a fixed distance. Such a random process has no memory by definition,
which implies that random fluctuations are uncorrelated and
incoherent in a time profile.
Earlier studies suspected that energy storage in solar flares 
accumulates as a linear function of time, which implies a
correlation between the waiting time and the energy released 
during two subsequent flares (Rosner and Vaiana 1978), but such
a predicted correlation was never found (Lu 1995; Crosby 1996;
Wheatland 2000b; Georgoulis et al.~2001).

A second mechanism is a nonlinear system, where the total dissipated
energy grows nonlinearly, the energy dissipation rate or event rate
is not constant, and the resulting waiting time distribution is
close to a power law distribution, $N(\tau) \propto \tau^{-\alpha}$.
The time evolution of an exponentially growing system is produced
by a coherent amplification mechanism, triggered by an instability
of a system. In the parlance of self-organized criticality systems,
an avalanche takes place that grows coherently over the duration
of an event. This could be a magnetic reconnection process that is
very common in solar and stellar flare physics. The coherence of
an exponentially growing system implies that there is a memory
effect over the duration of an event. 
The basic behavior of an avalanching system in terms
of next-neighbor interactions in a lattice grid has been simulated
extensively (Bak et al.~1987, 1988; Pruessner 2012). The critical diagnostic 
of coherent versus incoherent growth is quantified here with
a nonlinearity index $p$, from which we can obtain information
on what time scales a nonlinear system has memory, and whether
observed fluctuations are due to random noise or coherent time
structures with memory. It appears that the solar flare rate
exhibits memory from time scales of hours (during clustered
flares in an unstable active retion) to decades, driven by
the magnetic (Hale) solar cycle. The number of (coherent) 
detected time structures as a function of the time resolution
is shown in Fig.~5, which obeys the upper limit 
$n_{det} \le$ (1 year)/$n_{bin}$ (Fig.~5, dashed linestyle). 

A third mechanism is the limit-cycle behavior of a coupled
nonlinear system, which exhibits an oscillary pattern. 
An example is the solar cycle, which oscillates between
a poloidal and a toroidal global magnetic field. 
The oscillatory behavior is often accomplished by a
driving force and a feedback force that balance each other
in a quasi-equilibrium phase space, although the two
counteracting forces are delayed to each other by about a 
half period. There exist also linearly (strictly periodic)
oscillating mechanisms (e.g., pendulum, planetary resonances,
coronal loop kink-mode oscillations),
in contrast to the less regular nonlinear mechanisms.
Since oscillations occur over durations much longer
than a single period, we can attribute a memory time
scale at least over the duration of the observation,
or during a time interval with a constant oscillation 
period. In other words, the oscillation period is
the memorized piece of information.

\section{	Conclusions	}

We analyze the waiting times in the largest data sets of 
solar flare events, obtained from GOES soft X-ray light
curves, by using variable time resolutions from 2 days
to 4 decades. The automated flare detection algorithm
gathers over $3 \times 10^5$ events, from which we
identify $\approx 10^3$ events with significant
coherent growth characteristics. We define the
nonlinear growth phase (rise time) in terms of
a polynomial (as well as logistic) time evolution,
which allows us to discriminate linear random
events ($p \approx 1$) from nonlinear energy
dissipation events ($p \gapprox 2$). The memory
time of the Sun is essentially defined by time
structures with coherent growth, such as
exponential-growing solar flare clusters. 
We obtain the following results:

\begin{enumerate}
\item{On time resolutions of $T \approx$ 1.5 months to 1 year,
the most prevailing nonlinear time structure is the
solar dynamo, which can be considered as a nonlinear
system with limit-cycle behavior, with an average 
period of 11 years. The degree of nonlinearity is
found to be $p = 2.8 \pm 0.5$, averaged over four
solar cycles and various time resolutions.
The corresponding power law slope is predicted to be
$\alpha=2+1/p\approx 2.4$, which is close to 
the value $(\alpha=2.5)$ calculated for a 
sinusoidally oscillating flare rate (Nurhan et al.~2021).
The sinusoidal flare rate variability can be understood 
by the oscillatory solar dynamo, where the poloidal
and the toroidal magnetic field vary sinusoidally
in anti-phase.} 

\item{Solar flares are found not to occur in random
order, although the flare rate time profile appears
to consist of many randomly scattered fluctuations
of the flare rate. Instead, clusters of flares are 
found at time resolutions from 2 hours to 3 days, 
which represents
some memory over these time scales. It indicates that
coherent growth in the flare rate is nearly quadratic, 
with a mean nonlinearity index of $p=2.2\pm0.1$.}

\item{At intermediate time resolutions from 6 days to
23 days, the solar rotation occults some flare
clusters partially, causing a lower but still
significant nonlinearity index of $p=1.8\pm0.2$.}  

\item{The determination of the nonlinearity index $p$
allows us to predict the power law slope $\alpha$ of the
waiting time distribution, $\alpha = 2.0 + 1/p$,
predicting values in the range of $\alpha=2.4-2.5$,
which agrees with the observational values of
$\alpha=2.1-2.4$.}

\item{The nonlinearity index range of $p \gapprox 2$
is consistent with the exponentially-growing 
characteristic of avalanches governed by
self-organizing criticality.}  

\end{enumerate}

The main new result of this study is the demonstration
that the Sun reveals memory over a huge range of time
scales, from a few hours to several decades, rather
than producing flares in random order. This means that
nonlinear physical processes produce spatio-temporal
structures with coherent time evolutions, such as
instabilities with exponential growth and subsequent
decay. There are self-organized avalanche processes
with clustered (sympathetic) flare generation, which 
occurs at a higher hierarchical level than the 
avalanches of individual flare events.

\bigskip
{\sl Acknowledgements:}
Part of the work was supported by NASA contracts 
NNG04EA00C and NNG09FA40C. JRJ acknowledges 
NASA grant NNX16AQ87G.

%%%%%%%%%%%%%%%%%%%%%%% REFERENCES %%%%%%%%%%%%%%%%%%%%%%%% 
\section*{ References }  

\def\ref#1{\par\noindent\hangindent1cm {#1}} 

\ref{Aschwanden, M.J., Dennis, B.R., and Benz, A.O. 1998, ApJ 497, 972}
\ref{Aschwanden, M.J. and McTiernan, J.M. 2010, ApJ 717, 683}
\ref{Aschwanden, M.J. 2011 {\sl Self-Organized Criticality in 
	Astrophysics.  The Statistics of Nonlinear Processes 
	in the Universe}, Springer-Praxis: New York, 416p.}
\ref{Aschwanden, M.J. 2012a, AA 539:A2}
\ref{Aschwanden, M.J. 2012b, ApJ 757, 94}
\ref{Aschwanden, M.J. and Freeland, S.M. 2012, ApJ 754:112}
\ref{Aschwanden, M.J., Crosby, N., Dimitropoulouy, M., et al.~2016,
	SSRv 198:47}
\ref{Aschwanden, M.J. 2019a, ApJ 880, 105}
\ref{Aschwanden, M.J. 2019b, ApJ 887:57}
\ref{Aschwanden, M.J. and Dudok de Wit, T. 2021, ApJ 912:94}
\ref{Aschwanden, M.J., Johnson, J.R., and Nurhan, Y.I., 2021,
	GRL (subm.)} 
\ref{Bak, P., Tang, C., and Wiesenfeld, K. 1987,
        Physical Review Lett. 59/27, 381}
\ref{Bak, P., Tang, C., and Wiesenfeld, K. 1988,
        Physical Rev. A 38/1, 364}
\ref{Bauke, H. 2007, Eur.Phys.J.B. 58, 167}
\ref{Beltrami, E. 1987, {\sl Mathematics for dynamic modeling},
 	Academic Press, Inc: Boston}
\ref{Boffetta, G., Carbone, V., Giuliani, P., Veltri, P., 
	and Vulpiani, A. 1999, Phys.Rev.Lett. 83, 4662}
\ref{Bourouaine, S., Perez, J.C., Klein, K.C., Chen, C.H.K. 2020,
	ApJ 904, 308}
\ref{Cameron, R. and Schuessler, M. 2017, ApJ 843, 111}
\ref{Charbonneau, P. 2005, Living Reviews in Solar Physics 2, 2}
\ref{Consolini, G., Tozzi, R., and De Michelis P. 2009,
	A\&A 506, 1381}
\ref{Crosby, N.B. 1996, PhD Thesis, University Paris VII, 
	Meudon, Paris, 348p.}
\ref{Dudok de Wit, P., Krasnoselskikh, V.V., Bale, S.D., et al. 
	2020, ApJSS  246:39}
\ref{Eastwood, J.P., Wheatland, M.S., Hudson, H.S., et al.
 	2010, ApJ 708, L95}
\ref{Greco, A., Matthaeus, W.H., Servidio, S., Chuychai, P.
	and Dmitruk, P. 2009, ApJ 2, L111} 
\ref{Georgoulis,M.K., Vilmer,N., and Croby,N.B. 2001,
 	AA 367, 326}
\ref{Goldstein, M.L., Morris, S.A., and Yen, G.G. 2004,
        Eur.Phys.J.B. 41, 255}
\ref{Gorobets, A. and Messerotti, M. 2012, SoPh 281, 651}
\ref{Grigolini, P., Leddon,D., and Scafetta,N. 2002,
 	Phys.Rev.Lett E, 65/4. id. 046203}
\ref{Guidorzi,C., Dichiara, S., Frontera,F., et al. 
 	2015, ApJ 801, 57}
\ref{Hawley, S.L., Davenport, J.R.A., Kowalski, A.F., Wisniewski, 
	John P. et al. 2014, ApJ 797, 12H} 
\ref{Hudson, H.S. 2020, MNRAS 491, 4435}
\ref{Jackson, E.A. 1989, {\sl Perspectives of nonlinear 
 	dynamics (Lotka-Volterra equation}, p.283),
 	Cambridge University Press: Cambridge}
\ref{Leddon, D. 2001, eprint arXiv:cond-mat/0108062, 
	Dissertation Abstracts International, Volume: 63-12, 
	Section: B, page: 5891; 56 p.}
\ref{Lepreti, F., Carbone, V., and Veltri,P. 2001, ApJ 555, L133}
\ref{Li, C., Zhong, S.J., Wang, L., et al. 2014, ApJ 792, L26}
\ref{Li, C., Zhong, S.J., Xu, Z.G., et al. 2018, MNRAS 479 L139}
\ref{Lu, E.T. 1995, ApJ 447, 416}
\ref{May, R.M. 1974, {\sl Model Ecosystems}, Princeton}
\ref{Morales, L.F. and Santos, N.A. 2020, SoPh 295, 155}
\ref{Norman, J.P., Charbonneau, P., McIntosh, S.W., and Liu, H.
 	2001, ApJ 557, 891}
\ref{Nurhan, Y.I., Johnson, J.R., Homan, R., et al.~(2021), GRL, (subm.)}
\ref{Newman, M.E.J. 2005, Contemp.Phys. 46, 323}
\ref{Pruessner, G. 2012, {\sl Self-organised criticality. 
	Theory, models and characterisation},
 	Cambridge University Press: Cambridge}
\ref{Qin, G. and Wu, S.S. 2018, ApJ 869, 48}
\ref{Rosner,R., and Vaiana,G.S. 1978, ApJ 222, 1104}
\ref{Usoskin, I.G., Solanki, S.K., and Kovaltsov, G.A. 
	2017, AA 471, 301}
\ref{Wang, F.Y., Dai, Z.G., Yi, S.X., and Xi, S.Q. 2015, ApJS 216, 8}
\ref{Wang, H.N., Cui, Y.M., and He,H. 2009, Res.Astron.Astrophys. 9, 687}
\ref{Wang, J.S., Wang, F.Y., and Dai, Z.G. 2017, MNRAS 471 2517}
\ref{Wang, Y., Liu, L., Shen, C., et al. 2013, ApJ 763, L43}
\ref{Wanliss, J.A. and Eygand, J.M. 2007, GRL 34, 4107}
\ref{Watari, S. 1996a, Solar Phys 163, 371}
\ref{Watari, S. 1996b, Solar Phys 168, 413}
\ref{Wheatland, M.S., Sturrock,P.A., and McTiernan,J.M. 1998, 
	ApJ 509, 448}
\ref{Wheatland,M.S. 2000a, ApJ 536, L109}
\ref{Wheatland,M.S. 2000b, SoPh 191, 381}
\ref{Wheatland,M.S. 2000c, ApJ 532, 1209}
\ref{Wheatland, M.S. and Litvinenko, Y.E. 2002, SoPh 211, 255}
\ref{Wheatland, M.S. 2003, SoPh 214, 361}
\ref{Wheatland, M.S., and Craig, I.J.D. 2006, SoPh 238, 73}
\ref{Willis, J.C. and Yule, G.U. 1922, Nature 109, 177} 
\ref{Yeh, C.T., Ding, M., and Chen, P. 2005, ChJAA 5, 193}
\ref{Yi, S.X., Xi, S.Q., Yu, Hai et al. 2016, ApJS 224, 20}
%%%%%%%%%%%%%%%%%%%%%%% REFERENCES %%%%%%%%%%%%%%%%%%%%%%%% 

%%%%%%%%%%%%%%%%%%%%%%%%%%% TABLE 1 &&&&&&&&&&&&&&&&&&&&&&&&&&&&&&&&&
\begin{table}
\begin{center}
\normalsize
\caption{Nonlinearity index $p(\Delta t)$ 
of flare rate function $y(t)$ as a function of the time
resolution $\Delta t$.}
\medskip
\begin{tabular}{lcclll}
\hline
Time	   & Number of   & Number of & Nonlinearity & Nonlinearity & Nonlinear \\
resolution & time bins   & events    & index        & median       & system\\
$\Delta t$ &$n_{bin}/yr$ & $n_{ev}$  & $p$          & $p_{med}$	   & \\
\hline
\hline
1 yr	  & 1	& 1	& 3.51$\pm$0.08	& 3.51 & solar dynamo \\
6 months  & 2	& 4	& 2.54$\pm$0.26 & 0.26 & solar dynamo \\
3 months  & 4	& 4	& 2.83$\pm$1.70 & 2.79 & solar dynamo \\
1.5 month & 8	& 8	& 2.43$\pm$1.76 & 1.91 & solar dynamo \\
\hline
23 days	  & 16	& 16	& 1.63$\pm$1.40	& 1.18 & solar rotation\\
11 days	  & 32  & 18    & 1.81$\pm$1.54 & 1.47 & solar rotation\\
6 days    & 64  & 59	& 2.06$\pm$1.78 & 1.42 & solar rotation\\
\hline
3 days    & 128 & 70    & 2.25$\pm$1.91 & 1.63 & solar flares\\
1.4 days  & 256 & 91    & 2.12$\pm$1.61 & 1.45 & solar flares\\
17 hours  & 512 & 108   & 2.15$\pm$1.83 & 1.57 & solar flares\\
8 hours   & 1024 & 137  & 2.23$\pm$1.55 & 1.88 & solar flares\\
4 hours   & 2048 & 195  & 2.11$\pm$1.22 & 1.84 & solar flares\\
2 hours   & 4096 & 206  & 2.29$\pm$1.37 & 2.11 & solar flares\\
\hline
\end{tabular}
\end{center}
\end{table}

%%%%%%%%%%%%%%%%%%%%%%%%%%%%%%%% TABLE 2S %%%%%%%%%%%%%%%%%%%%%%%%%%%%%%%%

\begin{figure}		
\centerline{\includegraphics[width=1.0\textwidth]{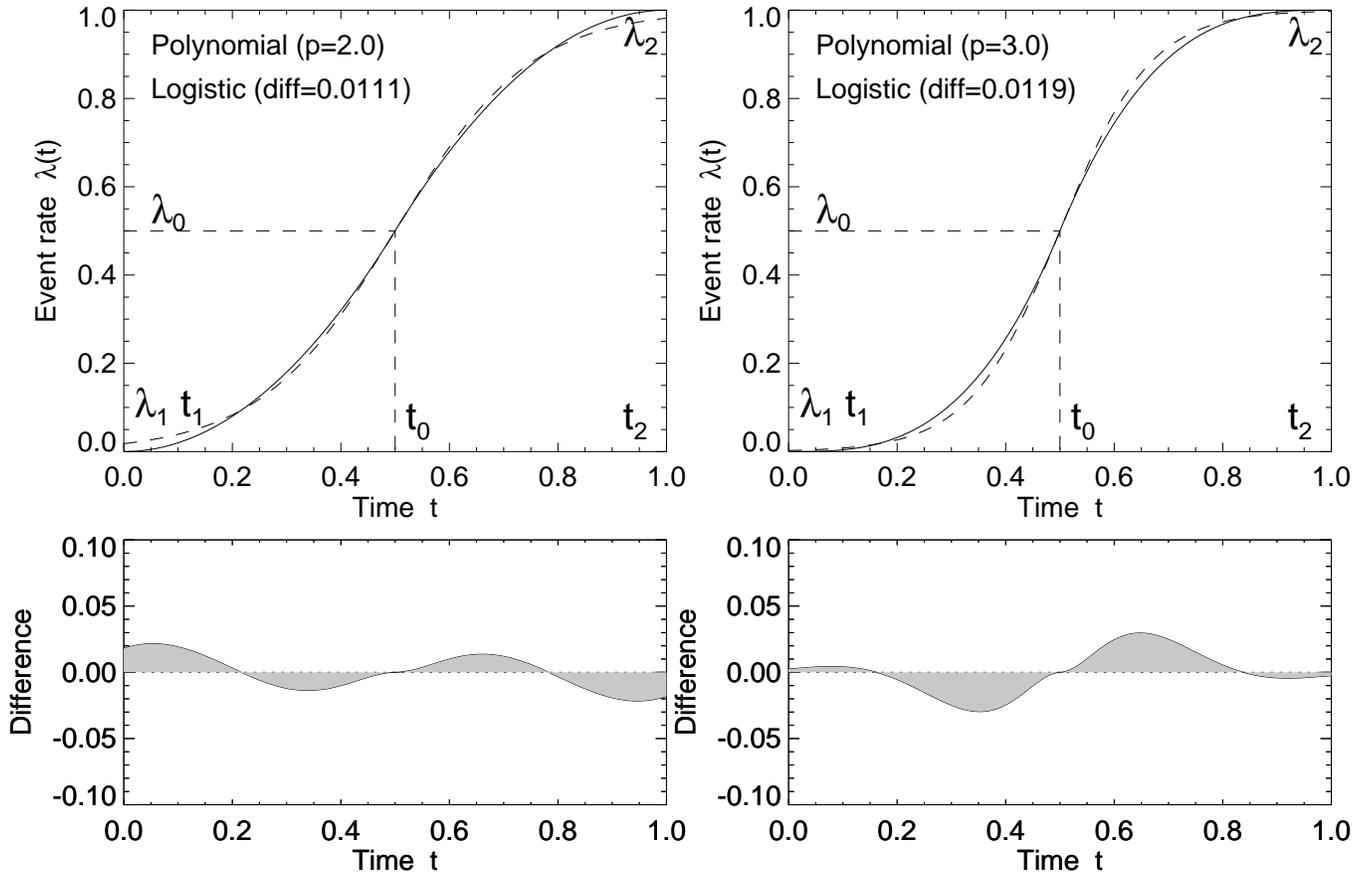}}
\caption{The time profile of a flare rate function $\lambda)(t)$
is modeled with polynomial functions (Eq.~6), here 
with a quadratic function, $\lambda_{pol}(t) \propto t^p$, with $p=2$
(left; solid curve), and with a cubic function, $\lambda_{pol}(t) \propto t^3$, 
with $p=3$ (right; solid curve).
Alternatively, the time profile $\lambda_{log}$ is modeled with the 
logistic equation (Eq.~8) (dashed curve),
consisting of an initial exponential rise phase with 
subsequent saturation. The rise time is defined within 
the time range of $[t_1, t_2]$. The logistic model is almost 
identical to the polynomial model, with a mean difference
of $|\lambda_{log}(t)-\lambda_{pol}(t)| \approx 1\% $.}
\end{figure}

\begin{figure}		
\centerline{\includegraphics[width=1.0\textwidth]{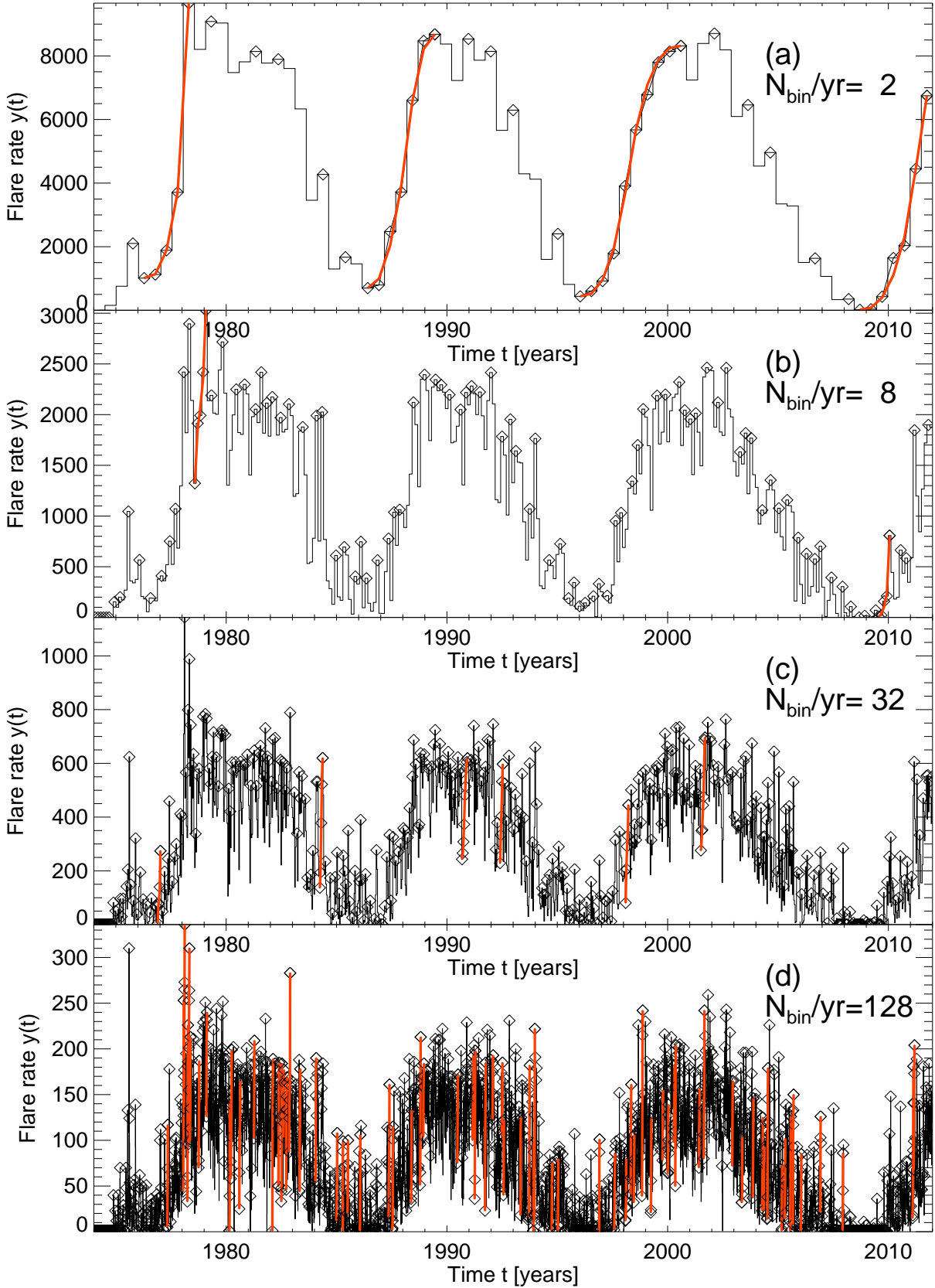}}
\caption{Time profile (histograms) of the flare rate $\Lambda(t)$ 
during the time range of 37 years (1974-2012) with 
four different time resolutions of 2 (a), 8 (b), 32 (c), and 
128 bins per year (d). The red curves represent best fits
of the polynomial flare rate function $\lambda(t)$. The 
local peaks of the histograms are marked with diamonds.}
\end{figure}

\begin{figure}		%BoundingBox: 0 0 581 400
\centerline{\includegraphics[width=1.0\textwidth]{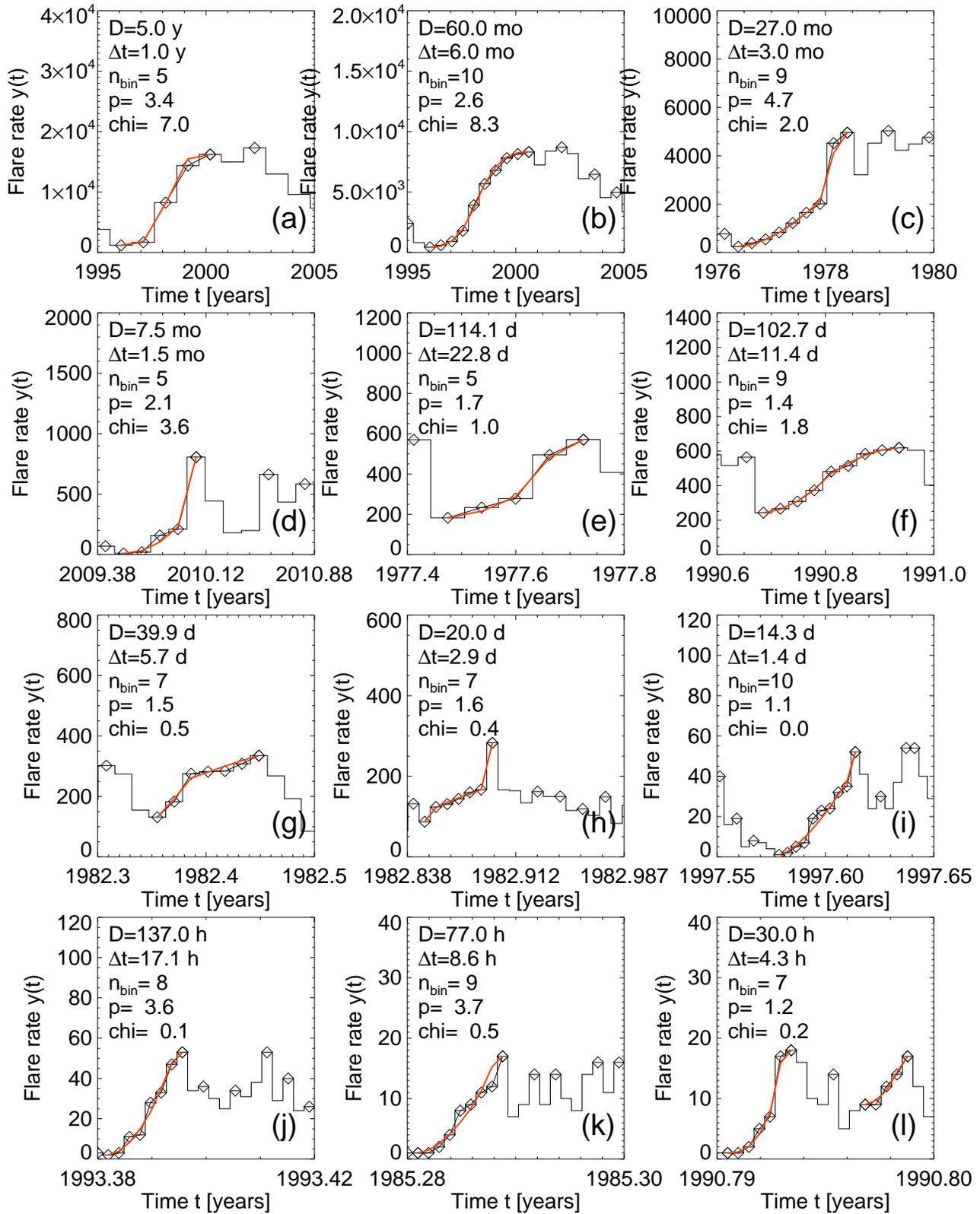}}
\caption{Fitted time profiles $\lambda(t)$ for 12 events with
increasing temporal solutions, from $\Delta t=1$ yr (panel a)
to $\Delta t=4.3$ hrs (panel l).}
\end{figure}

\begin{figure}		%BoundingBox: 0 0 581 400
\centerline{\includegraphics[width=1.0\textwidth]{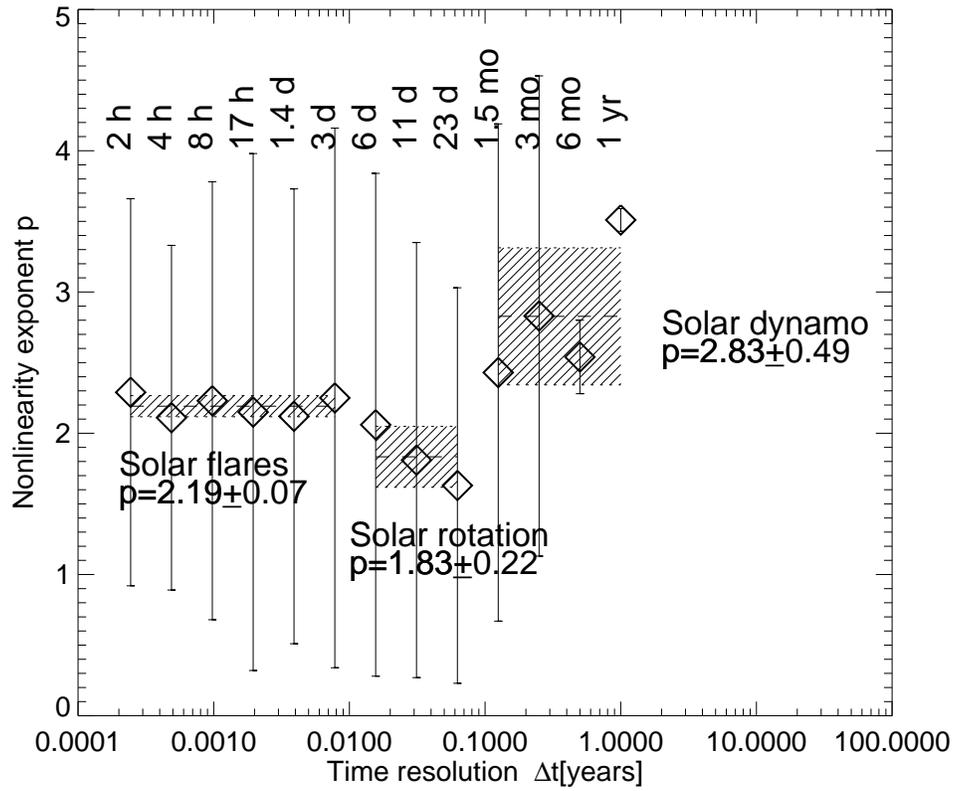}}
\caption{The nonlinearity index $p(\Delta t)$ as a function
of the time resolution $\Delta t$ is calculated
from an automated GOES flare detection algorithm. Note the triple
regimes of solar flares 
(in the time range of $\Delta t \approx$ 2 hrs - 3 days), 
the solar rotation 
(in the time range of $\Delta t \approx$ 6 - 45 days), 
and the solar dynamo 
(in the time range of $\Delta t \approx$ 1.5 months - 1 year).}
\end{figure}

\begin{figure}		%BoundingBox: 0 0 581 400
\centerline{\includegraphics[width=1.0\textwidth]{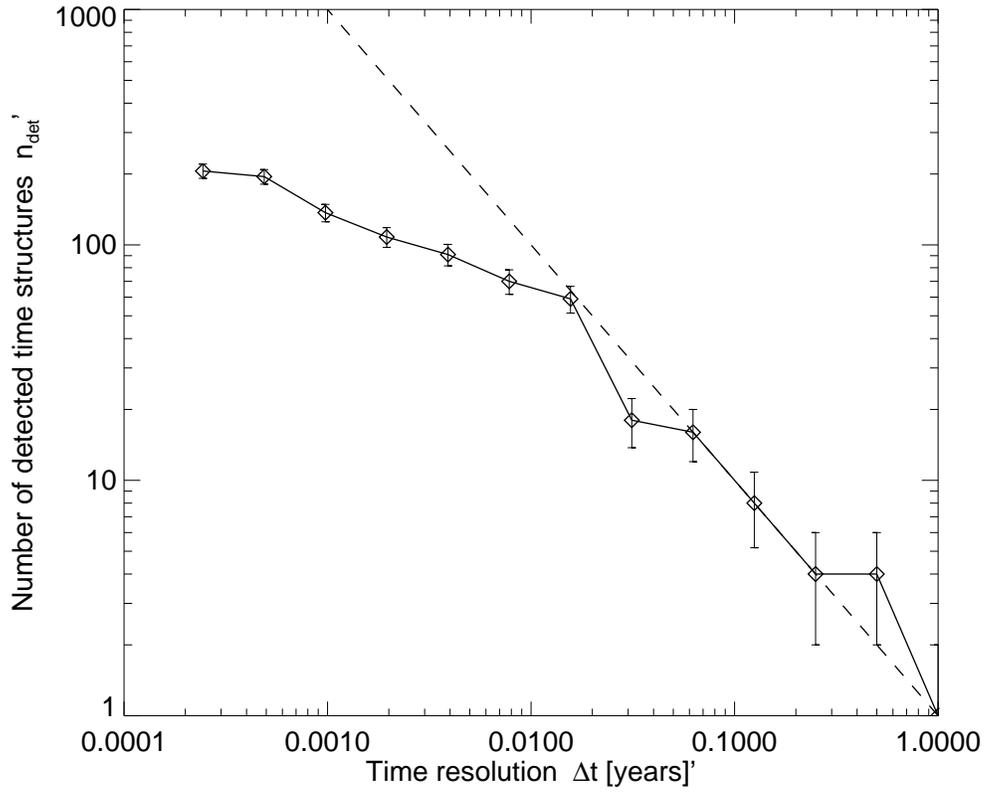}}
\caption{The number of detected time structures $n_{det}$ as
a function of the time resolution (thick linestyle with
error bars) and predicted upper limit (dashed linestyle).}
\end{figure}

\end{document}